\documentclass[journal]{IEEEtran}
\usepackage{cite}

\ifCLASSINFOpdf
  \usepackage[pdftex]{graphicx}
\else
\fi
\usepackage{algorithmic}

\usepackage{tabularx}
\usepackage{booktabs}
\usepackage{float}

\usepackage{float}

\usepackage{stfloats}
\usepackage[hyphens]{url}

\usepackage[usenames, dvipsnames]{color}
\usepackage[disable]{todonotes}
\definecolor{red}{rgb}{0,0,0}

\usepackage[absolute,overlay]{textpos}
\begin{document}
%
\title{Adaptive Music Composition for Games}
%
%
%

\author{Patrick Hutchings,
        Jon McCormack
\thanks{P. Hutchings and J. McCormack are with Monash University.}}

%
%

\markboth{IEEE Transactions on Games (preprint)}%
{Shell \MakeLowercase{\textit{et al.}}: Bare Demo of IEEEtran.cls for IEEE Journals}
%



\maketitle

%
\begin{textblock*}{\textwidth}(1.2cm,26.8cm) 
  \fbox{%
    \parbox{\textwidth}{\small%
        Preprint of: P.~Hutchings and J.~McCormack, `Adaptive Music Composition for Games', 
		IEEE Transactions on Games, July 2019, pp. 1-11, doi: 10.1109/TG.2019.2921979
          } }
\end{textblock*}

\begin{abstract}
The generation of music that adapts dynamically to content and actions has an important role in building more immersive, memorable and emotive game experiences.  To date, the development of adaptive music systems for video games is limited by both the nature of algorithms used for real-time music generation and the limited modelling of player action, game world context and emotion in current games.  We propose that these issues must be addressed in tandem for the quality and flexibility of adaptive game music to significantly improve.  Cognitive models of knowledge organisation and emotional affect are integrated with  multi-modal, multi-agent composition techniques to produce a novel Adaptive Music System (AMS).  The system is integrated into two stylistically distinct games.  Gamers reported an overall higher immersion and correlation of music with game-world concepts with the AMS than with the original game soundtracks in both games.
\end{abstract}



%
\IEEEpeerreviewmaketitle

\section{Introduction}

\IEEEPARstart{V}{ideo game} experiences are increasingly dynamic and player directed, incorporating user-generated content and branching decision trees that result in complex and emergent narratives \cite{louchart2015emergent}.  Player-directed events can be limitless in scope but their significance to the player may be similar or greater than those directly designed by the game's developers.  Game music, however continues to have far less structural dynamism than many other aspects of the gaming experience, limiting the effects that unique, context specific music scores can add to gameplay.  Why shouldn't unpredicted, emergent moments have unique musical identities? As the soundtrack is an inseparable facet of the shower scene in Hitchcock's `Psycho', context-specific music can help create memorable and powerful experiences that actively engage multiple senses.  

Significant, documented effects on memory \cite{boltz2004cognitive}, immersion \cite{sanders2010time} and emotion perception \cite{sloboda1991music} can be achieved by combining visual content and narrative events with music containing specific emotive qualities and associations.  \textcolor{red}{In games, music can also be assessed in terms of narrative fit and functional fit - how sound supports playing} \cite{ekman2014cognitive}. The current standard practice of constructing music for games by starting music tracks or layers with explicitly defined event triggers, allows for a range of expected situations to be scored with a high confidence of musical quality. However, this introduces extensive repetition and inability to produce unique musical moments for a range of unpredicted events. 

In this paper we present an adaptive music system (AMS) based on cognitive models of emotion and knowledge organisation in combination with a multi-agent algorithmic music composition and arranging system.  We propose that a missing, essential step in producing affective music scores for video games is a communicative model for relating events, contents and moods of game situations to the emotional language of music.  This would support the creation of unique musical moments that reference past situations and established relationships without excessive repetition.  Moreover, player-driven events and situations not anticipated by the game's developers are accommodated.

\begin{figure*}
\begin{center}
\includegraphics[width=0.9\textwidth]{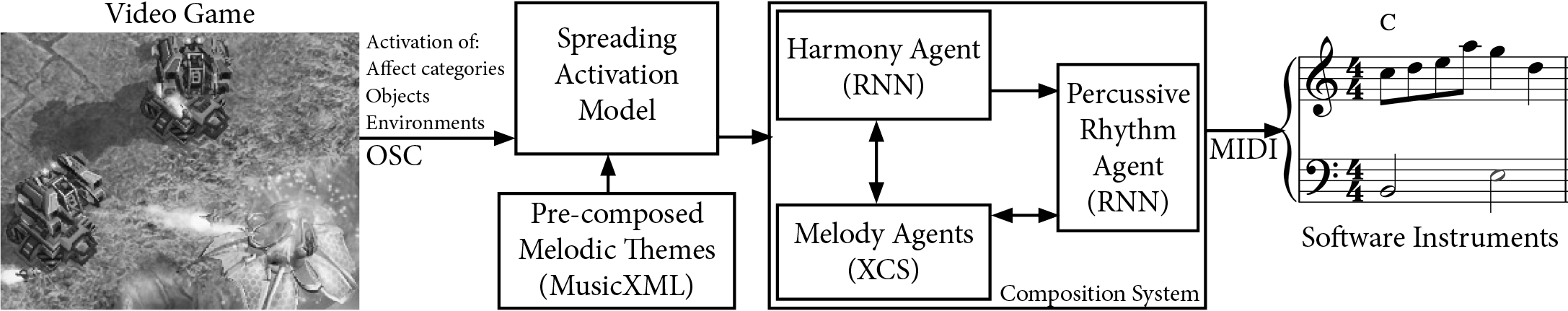}
\caption{Architecture of the Adaptive Music System (AMS) for modelling game-world context and generating context-specific music.}
\label{fig:architecture}
\end{center}
\end{figure*}

The \emph{spreading activation model} of semantic content organisation used in cognitive science research is adopted as a generalised model that combines explicit and inferred relationships between emotion and objects and environments in the game world.  Context descriptions generated by the model are fed to a multi-agent music composition system that combines recurrent neural networks with evolutionary classifiers to arrange and develop melodic contents sourced from a human composer (see Figure \ref{fig:architecture}).

\section{Related Work}

This research is primarily concerned with `Experience-Driven Procedural Content Generation', a term coined by Yannakakis and Togelius \cite{yannakakis2011experience} and applied to many aspects of game design \cite{shaker2016procedural}.  As such, the research process and methodology are driven by questions of the experience of music and games, individually and in combination.  We frame video games as virtual worlds and model game contents from the perspective of the gamer.  The virtual worlds of games include their own rules of interaction and behaviour that overlap and contrast with the physical world around us, and cognitive models of knowledge organisation, emotion and behaviour give clues about how these worlds are perceived and navigated.  This framing builds on current academic and commercial systems of generating music for games based on gameworld events, contents and player actions.

Music in most modern video games displays \emph{some} adaptive behaviour.  Dynamic layering of instrument parts has become particularly common in commercial games, where instantaneous environment changes, such as speed of movement, number of competing agents and player actions, add or remove sonic layers of the score.  Tracks can be triggered to play before or during a scripted event or modified with filters to communicate gameplay concepts such as player health.  Details of the music composition systems from two representative popular games, as revealed through creator interviews, show the state of generative music in typical, large-budget releases.

\textit{Red Dead Redemption} \cite{game:reddead} is set in a re-imagined wild American west and Mexico. The game is notable in its detailed layering of pre-recorded instrument tracks to accompany different gameplay events. To accommodate various recombinations of instrument tracks, most of the music sequences were written in the key of A minor and a tempo of 130 beats per minute ~\cite{reddead}. Triggers such as climbing on a horse, enemies appearing and time of day in the game world add instrumental layers to the score to build up energy, tension or mood.

\textit{No Man's Sky} \cite{game:nomanssky} utilises a generative music system and procedural content generation of game worlds.  Audio director Paul Weir has described the system as drawing from hundreds of small carefully isolated fragments of tracks that are recombined following a set of rules built around game-world events. Weir mentions the use of stochastic choices of melodic fragments but has explicitly denied the use of genetic algorithms or Markov chains \cite{Epstein}.

\textcolor{red}{Adaptive and generative music systems for games have an extensive history in academic research.  From 2007 to 2010 a number of books were written on the topic \cite{collins1975spreading,grimshaw2010game,farnell2007introduction} but over the last decade, translation from research to wide adoption in commercial games has been slow.} 

\textcolor{red}{Music systems have been developed as components of multi-functional proceedural content generation systems.  Extensions of the dynamic layering approach were implemented in \textit{Sonancia} \cite{lopes2015sonancia}, which triggers audio tracks by entering rooms generated with associated affect descriptions.}

Recent works in academia have aimed to introduce the use of emotion metrics to direct music composition in games, but typically with novel contributions in either composition algorithms or modelling context in games.  

\textcolor{red}{Eladhari et al. utilised a simple spreading activation model with emotion and mood nodes in the \textit{Mind Music} system that was used to inform musical arrangements for a single character in a video game \cite{eladhari2006soundtrack}, but like many adaptive music systems for video games presented in the academic literature \cite{lopes2015sonancia,hoover2015audioinspace,holtar2013audioverdrive,ibanez2017towards} there is no evaluation through user studies or predefined metrics of quality to judge the effectiveness of the implementation or underlying concepts.}

Recent work by Ib{\'a}{\~n}ez et al. \cite{ibanez2017towards} used an emotion model to mediate music scoring decisions based on game-dialogue mediated with primary, secondary and tertiary emotion activations of discrete emotion categories. The composition system used pre-recorded tracks that were triggered by the specific emotion description.  In contrast  Scirea et al. \cite{scirea2017affective} developed a novel music composition model, MetaCompose, that combines graph traversal-based chord sequence generation with an evolutionary system of melody generation.  Their system produces music for general affective states -- either negative or positive -- through the use of dissonance.

This research treats the modelling of relationships between elements of game-worlds and emotion as fundamental to music composition, as they are key aspects of the compositional process of expert human composers.  As the famed Hollywood film composer Howard Shore states: \textit{``I want to write and feel the drama. Music is essentially an emotional language, so you want to feel something from the relationships and build music based on those feelings.''} \cite{howardShore}.

\section{Algorithmic Techniques}
\todo[inline]{renamed section from `research background'}

Algorithmic music composition and arranging with computers has developed alongside, but mostly separately from video games.  While algorithms have become more advanced, most techniques lack the expressive control and musical consistency desired in a full composition system for use in commercial video games.  Algorithmic composition techniques exhibit a range of strengths and weaknesses, particularly in regards to consistency, transparency and flexibility.

Evolutionary systems have been popular in music generation systems, with a number of new evolutionary music systems developed in the 1990s and early 2000s in particular \cite{de2012special,miranda2007evolutionary}, allowing for optimised solutions for problems that cannot be solved efficiently or have changing, unpredictable dynamics.  Evolutionary music systems often utilise expert knowledge of music composition \textcolor{red}{in the design of} fitness functions and can produce complexity from a small set of rules. These properties result in greater flexibility, but with reduced predictability or consistency.

\todo[inline]{Moved XCS description from architecture section}

Wilson's \emph{XCS} is an eXtended learning Classifier System that uses an evolutionary algorithm to evolve rules, or classifiers, that describe actions to be taken in a given situation \cite{wilson1995classifier}.  In XCS, classifiers are given a fitness score based on the difference between the reward expected for a particular action compared to the reward received.  By adapting to minimise predictive error rather than just maximising reward, classifiers that are only rewarded well in specific and rare contexts can remain in the population.  This helps prevent convergence to local minima -- particularly important in the highly dynamic context of music.  XCS has been successfully used in creative applications to generate visuals and sound for dynamic artificial environments \cite{mccormack2007artificial} to take advantage of these properties.

Neural networks have been used for music generation since the early 1990s \cite{mozer1994neural,papadopoulos1999ai,mccormack2009generative} but have recently grown in popularity due to improvements in model design and growth of parallel compute power in PCs with graphical processing units (GPUs).  Surprisingly musical results have come out of applying NLP neural network models to music sequence generation \cite{eck2002first,sturm2016music}. These models have proven effective at structuring temporal events, including monophonic melody and chord sequence generation, but suffer from a lack of consistency, especially in polyphonic composition.  Deep learning with music audio recordings is a growing area of analysis and content generation \cite{van2016wavenet} but is currently too resource intensive to be used for real-time generation.

\section{The Experience of Interaction}
\todo[inline]{Experience, knowledge and emotion sections merged}

Most games are interactive and infer some intention or aspect of the gamer experience through assumptions of how the game action might be affecting the gamer.  Such assumptions are informed by internal models that game designers have established about how people experience interactive content and can be formalised into usable models.  Established methodologies for analysing experience have lead to demonstrated measurable psychological effects when interacting with virtual environments.

Sundar et al. \cite{sundar2015toward} outlined an agency based theory of interactive media and found empirical evidence suggesting that agency to change interactive media has strong psychological effects connected to identity, sense of control and persuasiveness of messages. The freedoms of the environment result in behavioural change. The user is a source of experience, knowledge and control in the virtual world, not just an observer.

Ryan et al. \cite{ryan2015open} listed four key challenges for interactive emergent narrative: \textit{modular content}, \textit{compositional representational strategies}, \textit{story recognition}, and \textit{story support}.  These challenges also affect dynamic music composition and can all be assisted by effective knowledge organisation.  By organising game content and events as units of knowledge connected by their shared innate properties and by their contextual role in the player's experience, new game states can be analysed as a web of dynamic contents and relationships.

\subsection{Knowledge Organisation}

Models of knowledge organisation aim to represent knowledge structures.  Buckley \cite{buckley2006theoretical} outlined properties of knowledge structures within the context of games:  ``\textit{... knowledge structures (a) develop from experience; (b) influence perception; (c) can become automated; (d) can contain affective states, behavioural scripts, and beliefs; and (e) are used to guide interpretations and responses to the environment.}''.

The spreading activation model \cite{collins1975spreading} is a knowledge organisation model that was first developed and validated \cite{lorch1982priming} to analyse semantic association \cite{battig1969category}, but has also been used as a model for other cognitive systems including imagery and emotion \cite{carr1982words}, supporting its suitability for use in video games.


The model is constructed as a graph of concept nodes that are connected by weighted edges representing the strength of the association between the concepts.  When a person thinks of a concept, concepts connected to it are activated to a degree proportional to the weight of the connecting edge.  Activation spreads as a function of the number of mediating edges and their weights. The spreading activation model conforms to the observed phenomena where words are recalled more quickly when they are primed with a similar word directly beforehand. However, it suffers as a predictive tool because it posits that the model network structure for each person is unique, so without mapping out an individual's associations it isn't possible to accurately predict how activation will spread.

Spreading activation models don't require logical structuring of concepts into classes or defining features, making it possible to add content based on context rather than structure.  For example, if a player builds a house out of blocks in \textit{Minecraft}, it does not need to be identified as a house.  Instead, its position in the graph could be inferred by time characters spend near it, activities carried out around it, or known objects stored inside it.  Heuristics based on known mechanisms of association can be used to turn observations into associations of concepts.  For example, co-presentation of words or objects can create associations through which spreading activation has been observed \cite{ratcliff1978priming}.

While the spreading activation model was originally applied to semantic contents, it has been successfully used to model activation of non-semantic concepts, including emotion \cite{bower1981mood}.  

\subsection{Emotion}

Despite a general awareness of the emotive quality that video games can possess \cite{sykes2003affective}, it is uncommon for games to include any formal metrics of the intended emotional effect of the game, or inferred emotional state of the gamer. Yet emotion perception plays a pivotal role in composing music \textcolor{red}{for games \cite{cunningham2011emotion}}, so having a suitable model for representing emotion is critical for any responsive composition system. 

The challenge of implementing an emotion model in a game to assist in directing music scoring decisions is made greater by fundamental issues of emotion modelling itself:  there is no universally accepted model of human emotion among cognitive science researchers.  Generalised emotion models can be overly complex or lacking in expressive detail and domain specific models lack validation in other domains.  Several music-specific models and measures of emotion have been produced \cite{gabrielsson2001emotion,juslin2004expression,zentner2008emotions}.  The Geneva Emotions in Music Scales (GEMS) \cite{zentner2008emotions} were developed for describing emotions that people feel and perceive when listening to music, but have been shown to have reduced consistency with music styles outside of the western classical tradition \cite{lykartsis2013emotionality}.  Music-specific emotion models lack general applicability in describing a whole game context as they are based on a music listening-only modality.

\todo[inline]{Removed example spreading activation graph}

A spreading activation network that contains vertices representing emotions can be used to model how recalling certain objects can stimulate affect. In video games this can be exploited by presenting particular objects to try and trigger an affective response or by trying to affect the user in a way that activates their memory of specific objects. An object that the user associates with threat, such as a fierce enemy, could not only activate an affect of threat for the player but also other threatening objects in their memory through mediated activation.

\section{Emotion Perception Experiment}

For our system, a basic emotion model that uses five affect categories of \textit{happiness}, \textit{fear}, \textit{anger}, \textit{tenderness} and \textit{sadness} was adopted based on Juslin's observation of these being the most consistently used labels in describing music across multiple music listener studies \cite{juslin2013does}, as well as commonly appearing in a range of basic emotion models.  \textcolor{red}{Unlike the frequently utilised Russell's Circumplex model of affect \cite{russell1980circumplex}, this model allows for representations of mixed concurrent emotions which is a known affective property of music \cite{hunter2008mixed}.}

\textit{Excitement} was added as a category following consultation with professional game composers that revealed excitement to be an important aspect of emotion for scoring video games not covered in the basic affect categories.  These categories are not intended to represent a complete cognitive model, rather a targeted selection of terms that balance expressive range with generalisability within the context of scoring music for video games.  Although broken down into discrete categories, their use in a spreading activation model requires an activation scale, so can be understood as a six dimensional model of emotion \textcolor{red}{removing issues of reduced resolution found in standard discrete models, highlighted in Eerola and Vuoski's study on discrete and dimensional emotion models in music \cite{eerola2011comparison}.}  

A listener study was conducted to inform the AMS presented and evaluated in this paper.  The study was designed to look for correlations between music properties and perception of emotion in listeners over a range of compositions with different instrumentation and styles.  It was shared publicly on a website and completed by 134 participants.

Thirty original compositions were created, in jazz, rock, electronic and classical music styles with diverse instrumentations, each 20-40 seconds in length.  The compositions were manually notated and performed using sampled and synthesised instruments in \textit{Sibelius} and \textit{Logic Pro} software. 

\todo[inline]{removed screenshot figure of listener study}
Participants were asked to listen to two tracks and identify the track in which they perceived the higher amount of one emotion from the six discrete affect categories, presented as text in the middle of the interface with play buttons for each track.  \textcolor{red}{Pairwise assessment is an established approach for ranking music tracks by emotional properties \cite{madsen2012predictive}.} When both tracks were listened to in their entirety, participants could respond by clicking one button: Track A, Track B or Draw. Track A was selected randomly from a pool of tracks that had not been played in the last 10 comparisons.

The Glicko-2 chess rating system \cite{glickman2012example} was adopted for ranking tracks in each of the six affect categories.  With Glicko-2 ranking, players that are highly consistent have a low rating volatility and are less likely to move up or down in rank due to the results of a single match.  Each track had a rating, rating deviation and rating volatility, initialised with the default Glicko-2 values of 1500, 350 and 0.06 respectively.  The system adjusts the ratings, deviation and volatility of tracks after they have been compared with each other and ranks them by their rating.

Features of the note by note symbolic representations of the tracks were analysed for correlation with rankings in each category, summarised in Appendix Table \ref{tab:emotionTable}.

Analysis of track rankings and musical content of each track revealed a range of music properties (see Appendix Table \ref{tab:pitchEmotion}) correlate with perception of specific emotions.  These correlations were used in the design of the AMS to help it produce mood-appropriate music.

\section{AMS Architecture}

An AMS for scoring video games with context-specific, affective music was developed and integrated into two stylistically distinct games.  For testing with different, pre-built games, the AMS was developed as a stand-alone package that receives messages from video games in real-time to generate a model of the game-state and output music. The system consists of three key components: 

\begin{enumerate}

\item A spreading activation model of game context;
\item The six category model of emotion for describing game context;
\item An adaptive system for composing and arranging expert composed music fragments using these two models.

\end{enumerate}

\subsection{Spreading Activation}\label{spreading activation}

A model of spreading activation was implemented as a weighted, undirected graph $G = (V,E)$,  where $V$ is a set of vertices and $E$ a set of edges, using the Python library \textit{NetworkX}.

\begin{figure}[H]
\includegraphics[width=0.48\textwidth]{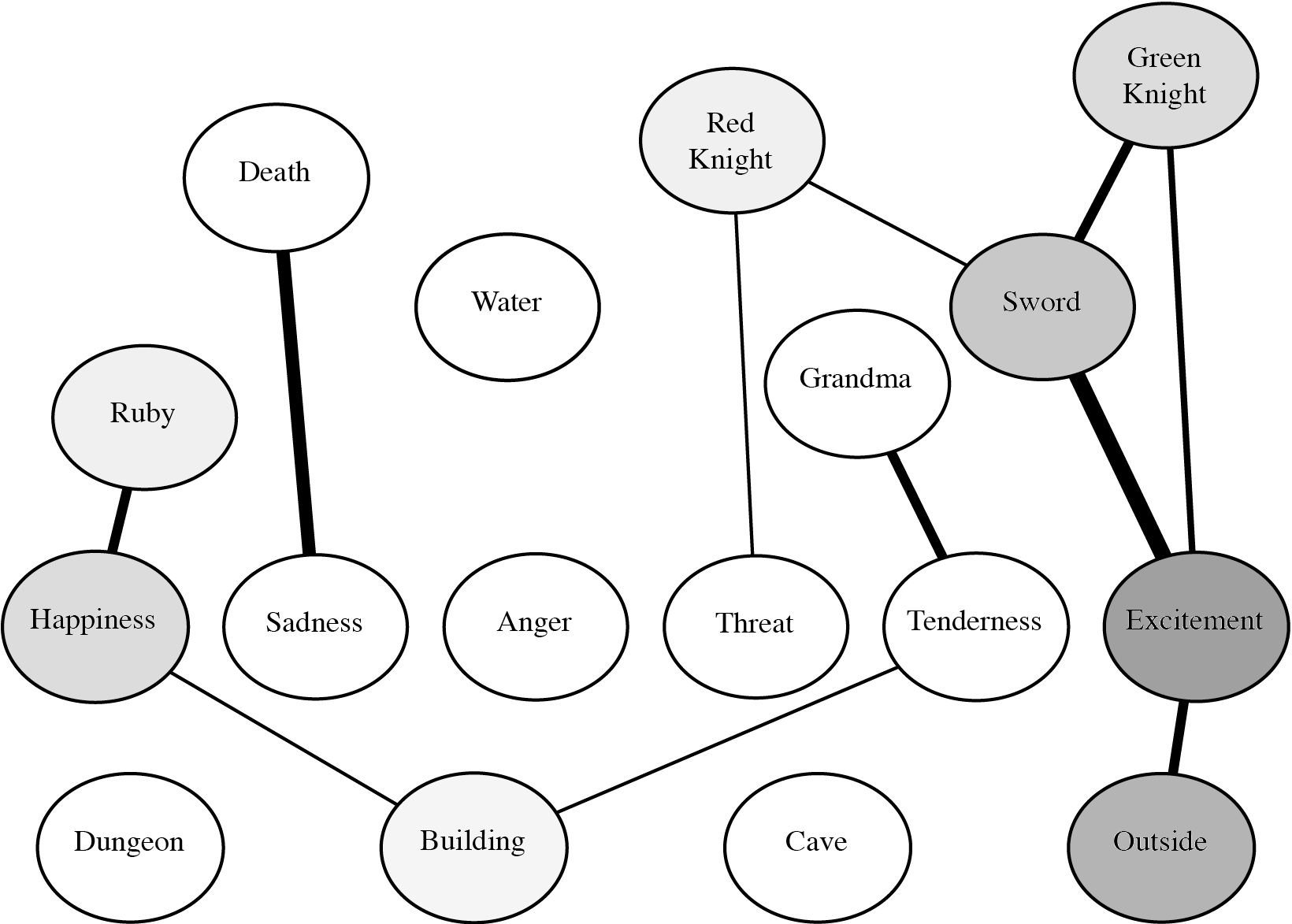}
\caption{Visualisation of spreading activation model during testing with a role-playing video game.  Vertex activation is represented by greyscale shading, linearly ramping from white to dark grey.  Edge weights were represented by edge width.}
\label{fig:visualisation}
\end{figure}

 Each vertex represents a concept or affect, with the weight of the vertex $w_V : V \Rightarrow A$ (for activation $A \in {\rm I\!R}$) representing activation between the values 0 and 100.  Normalised edge weights $w_E : E \Rightarrow C$ (for association strength $C \in {\rm I\!R}$) represent the level of association between vertices and are used to facilitate the spreading of activation.  Three node types are used to represent affect (A), objects (O) and environments (N) such that $V = A \cup O \cup N$.  The graph is not necessarily connected and edges never form between affect vertices.
 
\begin{figure}[h]
\begin{center}
\includegraphics[width=0.38\textwidth]{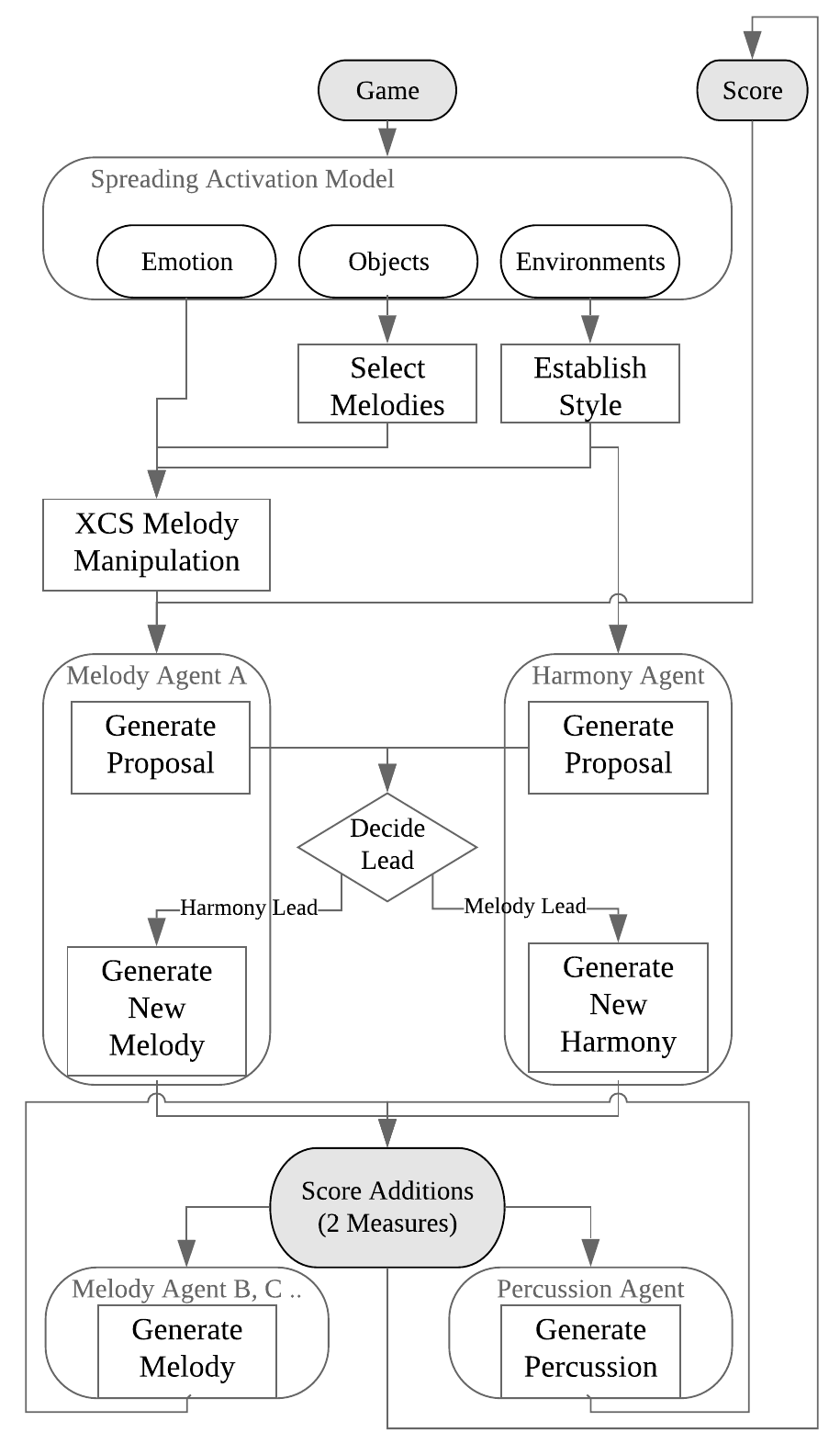}
\caption{\textcolor{red}{System Diagram for the AMS.}}
\label{fig:overview}
\end{center}
\end{figure}

A new instance of the graph with the six affect categories of \textit{sadness}, \textit{happiness}, \textit{threat}, \textit{anger}, \textit{tenderness} and \textit{excitement} is spawned whenever a new game is started.  The model is used to represent game context in real-time, regular messages from the game are used to update the graph.  These updates can add new vertices, new edges, or update edge weights.  An OSC \cite{wright2005open} client was added to the game engines to communicate the game state as lists of content activations and relationships for this purpose.

Every 30ms, the list of messages received by the OSC server is used to update the activation values of vertices in the graph.  If a message showing activation of a concept is received and the concept does not yet exist in the graph, a new vertex representing that concept is created.  At each update of the graph, edges from any vertex with activation above 0 are traversed and connected vertices are activated proportionally to the connecting edge weights.  For example, if vertex A has an activation of 50 and is connected to vertex B with an edge of weight 0.25 then vertex B would be activated to a minimum value of $50 \times 0.25 = 12.5$.  If vertex B is already activated above the this value, then no change is made to its activation.

Determining activation of concepts and affect categories from gameplay depends on the mechanics of the game, and would ideally be a consideration at all stages of game development.

Edges are formed between vertices using one of two methods.  Inferred edges are developed from co-activation of concept vertices.  When two concepts have an activation above 50 at the same time, an edge is formed between them.  Edges can also be assigned from game messages, allowing game creators to explicitly define associations.  For example, an edge between the vertices for the character `Grandma' and affect category `Tenderness' can be assigned by the game creators and communicated in a single message.

Activation of concepts, and edge weights, fade over time.  Fading rate is determined by the game length, number of concepts or game style as desired.  For the games tested, edge weights would not fade when the association was explicitly defined and inferred associations would fade at a rate of 0.01 per second, while vertex activations faded at a rate of 0.1 per second.

Melodic themes were precomposed for a subset of known objects in the game.  These themes were implemented as properties of object vertices (Eqn. \ref{themeEq}). 

\begin{equation} \label{themeEq}
v_t = \textit{theme}_{i}\   \forall \ v \in O
\end{equation}

At every frame, the knowledge graph is exposed to the music composition system and new music is generated to accompany gameplay.

\subsection{Music Composition}

A multi-agent composition system (Figure \ref{fig:overview}) was developed based on observations of group dynamics of improvising musicians and the relative strengths and weakness of different algorithmic composition techniques for distinct roles in the composition process.  The agent architecture was developed using a design science methodology \cite{hevner2007three} with iterative design cycles. A combination of music theory, including descriptive `rules' of harmonic tension and frequently used melody manipulation techniques, and calibration based on aural assessment of real-world performance guided these cycles.  The agents were assessed through a series of user studies, which provided feedback to support or reject more general aspects of approaches used.  

\subsubsection{Multi-agent Dynamics}

Co-agency is the concept of having individual agents that work towards a common goal but the agency to decide how to work towards that goal.  It is observed in team sports, where players rely on their own skills, knowledge and situational awareness to decide what actions to take, but work as a team to win the game.  Some teams have a `game plan' which give soft rules for how co-agency will be facilitated, but still leave room for independent thought and action.  Co-agency is also evident in improvising music ensembles \cite{sawyer2006group}.  Musicians add their own creative input to create original compositions as a group, using a starting melody, chord progression or rhythmic feel as their `game plan'.

\todo[inline]{Removed bit on codependency of melody and harmony.}

In the AMS, multiple software agents are used to generate musical arrangements.  Agents do not represent performers, instead roles that a player or multiple players in an improving group may take.  Agents were designed to have either \textit{Harmony}, \textit{Melody} or \textit{Percussive Rhythm} roles for developing polyphonic compositions with a mixture of percussive and pitched instruments.  Agents generate two measures worth of content at a time.

\subsubsection{Harmony Agent}

The harmony agent utilised a RNN model designed as an extension to the tested harmony system presented by Hutchings and McCormack \cite{hutchings2017using}.  The RNN contains two hidden layers, and an extensive hyper parameter search lead to the use of gated recurrent units (GRU) with 192 units per layer.  

Chord symbols were extracted from 1800 folk compositions from the Nottingham Music Database found at http://abc.sourceforge.net/NMD/, 2000 popular songs in rock, electronic, rhythm and blues and pop styles from the \textit{Wikifonia} database (no longer publicly available) and 2986 jazz standards from collection of the `Real Book' series of jazz standards \cite{realbook}. To aid the learning of chord patterns by style and encode time signature information into the training data, barlines were replaced with word tokens representing the style of music, either \textit{pop}, \textit{rock}, \textit{jazz} or \textit{folk}.

Dynamic unrolling produced significantly faster training times, and a lower perplexity was achieved through the implementation of peep-hole mechanisms.  Encoding of chord symbol tokens aided training, with best results achieved using 100 dimension encoding.  The final layer of the network utilised softmax activation to give a likelihood value for any of the chord symbols in the chord dictionary to come next for a given input sequence.  This allows the harmony agent to find likely next chords, add them to the chord progression of a composition and feed them back into itself to further generate new chords for a composition.

The output of the softmax layer is used as an agent confidence metric as used by Hutchings and McCormack \cite{hutchings2017using}.  The confidence of the agent is compared with other agents in the AMS to establish a `leader' agent at any given time.  When the harmony agent has a higher confidence score than the first melody agent, the harmony agent becomes the leader and the melody agent searches for a melody that is harmonically suitable.  If the harmony agent confidence is lower, less likely chords are progressively considered if needed to harmonically suit the melody.

The harmony agent does not add any notes to the game score, instead it produces a harmonic context which is used by multiple melody agents to write phrases for all the pitched instruments in the score.

\begin{table}[]
\begin{center}
\caption{ \textcolor{red}{Harmonic context representation as a 2d matrix for an example chord progression.}}
\label{tab:harmonicContext}
\begin{tabular}{l|llll|lllll}
	\ & \multicolumn{4}{l|}{C7} &  \multicolumn{4}{l}{E7} \\ \hline
	B & 0.3 & 0.3 & 0.3 & 0.3 & 0.8 & 0.8 & 0.8 & 0.8 \\ \hline
	A\# & 0.8 & 0.8 & 0.8 & 0.8 & 0.5 & 0.5 & 0.5 & 0.5 \\ \hline
	A & 0.3 & 0.3 & 0.3 & 0.3 & 0.3 & 0.3 & 0.3 & 0.3 \\ \hline
	G\# & 0.3 & 0.3 & 0.3 & 0.3 & 0.8 & 0.8 & 0.8 & 0.8 \\ \hline
	G & 0.8 & 0.8 & 0.8 & 0.8 & 0.5 & 0.5 & 0.5 & 0.5 \\ \hline
	F\# & 0.3 & 0.3 & 0.3 & 0.3 & 0.3 & 0.3 & 0.3 & 0.3 \\ \hline
	F & 0.3 & 0.3 & 0.3 & 0.3 & 0.3 & 0.3 & 0.3 & 0.3 \\ \hline
	E & 0.8 & 0.8 & 0.8 & 0.8 & 1.0 & 1.0 & 1.0 & 1.0 \\ \hline
	D\# & 0.3 & 0.3 & 0.3 & 0.3 & 0.3 & 0.3 & 0.3 & 0.3 \\ \hline
	D & 0.3 & 0.3 & 0.3 & 0.3 & 0.8 & 0.8 & 0.8 & 0.8 \\ \hline
	C\# & 0.3 & 0.3 & 0.3 & 0.3 & 0.3 & 0.3 & 0.3 & 0.3 \\ \hline
	C & 1.0 & 1.0 & 1.0 & 1.0 & 0.5 & 0.5 & 0.5 & 0.5 \\ \hline
\end{tabular}
\end{center}

\end{table}

The harmonic context \textcolor{red}{is a 2D matrix of values that represent which pitches most strongly signify a chord and key over the course of a musical measure \ref{tab:harmonicContext}.  For example, a row representing the pitch `C' would be populated with high values when the token output of the harmony agent is a C major chord.  In this situation the row representing the pitch `E flat' would have low values as hearing an E flat would suggest an E minor chord.  These values are formulated as resource scores that the melody agent can use to decide which combination of melodic notes will best reflect the chord and key.}  The matrix has 12 rows representing the 12 tones in the western equal temperament tuning system, from `C' to `B', and the columns represent subdivisions of beats across four measures.  When the harmony agent generates chords for two measures the matrix is populated with values that represent the available harmonic resource for the melody agents to use.  

Root tones are assigned a resource value of 1.0 and chord tones a value of 0.8.  All other values carry over from the previous measure to help establish tonality and are clamped to a range of 0-0.5 to prioritise chord tones.  The first measure is initialised with all values set to 0.3. 

The harmonic fitness score, $H$, is given by the average resource value at each cell \textcolor{red}{$K_i$} a note from the melody fragment inhabits:

\begin{equation}
\label{harmonicFitness}
H = \frac{1}{L}\sum_{i=1}^{L} f(K_i)
\end{equation}

\subsubsection{Melody Agents}

At any point of gameplay, the knowledge graph represents the activation of emotion, object and environment concepts.  The melodic theme assigned to the highest activated object is selected for use by melody agents when melodic content is needed in the score.  A melody agent exists for each instrument defined in the score by the human composer and each agent adds melodic content every two measures. 

XCS is used to evolve rules for modifying pre-composed melodic fragments using melody operators as actions. \textcolor{red}{It is used to reduce the search space of melody operations to support real-time responsiveness of the system.  Melody fragments of one to four measures in length are composed by expert composers as themes for different concepts in the game.} A list of eight melody operators were used based on their prevalent use in music composition: Reverse, Diminish, Augment, Invert, Reverse-Diminish, Reverse-Augment, Invert-Diminish and Invert-Augment.  \textcolor{red}{Reversal reverses the order of notes, augmentation and diminution extend and shorten the length of each note element-wise, respectively by a constant factor and inversion inverts the pitch steps between notes. The activation of each affect category is represented with a 2 bit binary number, to show concept activations of 0-24, 25-49, 50-74 or 75-100 (see Section \ref{spreading activation})}.  A 6-bit unique \emph{theme id} is appended to each of the emotion binary representations to create an 18-bit string representation of the environment.

Calculations of reward can be modified based on experimentation or adjusted to the taste of the composer.  The equations implemented as the default rewards were established using the results of the emotion survey (see Table \ref{tab:pitchEmotion}) to determine negative or positive correlation and then calibrated by ear (\textcolor{red}{by the lead author, who has professional experience composing for games}), to find reasonable bounding ranges and coefficients for each property.  Notes per second ($n_s$), mean pitch interval ($\bar{n}$) and ratio of diatonic to non-diatonic notes (d) in each phrase were used for the calculation of rewards:

\begin{algorithmic} \label{reward}
   \STATE $R_e = 0.2 - |(e/500) -(n_s-0.5)/25|$
   \STATE $R_h = 0.2 - | h - d|$
   \STATE $R_s = |(s/500) -(n_s-0.5)/25|$
   \STATE $R_{te} = |(te/500) -(n_s-0.5)/25|$
   \STATE $R_{th} = 0.2-|(th/500) -(\bar{p}/6)/5|$
   \STATE $R = R_e + R_h + R_s + R_{te} + R_{th}$
\end{algorithmic}

Only melodic fragments resulting from rules with rewards estimated above a threshold value ($R>0.6$ by default) are considered and the resulting modified melody fragments are put through an exhaustive search of transpositions and temporal shifts to find a suitable fit in the score based on the harmonic context.

If a suitable score is achieved, the melody fragment is placed into the music score and modifications are made to the harmonic context.  Notes consume all resources from the context where they are added, and consume half of the resources of tones directly above and below and a diminished fifth away.  By consuming resources from pitches with a high level of tension, notes with unpleasant clashes are avoided by other melody agents.

This process is repeated for each melody agent every two measures.  The first melody agent produces the melody line with the highest pitch, the second melody agent produces the lowest pitch melodic lines and subsequent melody agents gradually get higher above the lowest melody line until the highest melody line is reached.  This follows common harmonisation and counter-point composition techniques.

The maximum range, $M_r$, in semi-tones, between the highest note of the first melody agent and the lowest note of the second melody agent is determined by the number of melody agents in the system ($N$) and the style of music (see Eqn.~\ref{maximumRange}).  A `Style range-factor', $S_r$, was set to default values of 1.0, 0.8 and 0.7 respectively for jazz, pop and folk music styles, but can be adjusted by composers as desired.

\begin{equation}\label{maximumRange}
M_r = 12 \times S_r \times N
\end{equation}

By setting a minimum harmonic fitness threshold it is possible that not all agents will be used at all points of time.  This is good orchestration practice in any composition, and is often observed in group improvisations, where musicians will wait until they feel they have something to contribute to the composition before playing.

The harmony agent can generate chord progressions taking style labels as input (jazz, folk, rock and pop).  In contrast, the melody agent generates musical suggestions and \textcolor{red}{rates them using explicit metrics designed to represent different styles}.

There are many composition factors that influence style, including instrumentation, rhythmic emphasis, phrasing, harmonic complexity and melodic shape.  Instrumentation and performance components such as swing can be managed by higher level processes of the AMS but the melody agents themselves should consider melodic shape, harmony and rhythmic qualities when adding to the composition.  In this system a simple style score ($P$) is introduced that rates rhythmic density emphasis based on style using ad-hoc parameters tuned by ear. Parameters of notes per beat ($n_b$) and binary value for phrase starting off the beat (1 = true, 0 = false) represented by $o_b$. For jazz, $P = |1 - n_b|+ o_b$.  For rock and pop, $P = |(1/N) - n_b| $.  For folk, $P = |1 - n_b|$.

A final melody fitness score, $M$, is calculated as the sum of the style and harmonic fitness scores, i.e.~$M = H + P$.

For concepts that do not have a pre-composed theme, a theme is evolved when the first edge is connected to its vertex.  Themes from the two closest object vertices, as calculated using Dijkstra's algorithm \cite{dijkstra1959note}, are used to produce a pool of parents, with manipulations used by the XCS classifiers applied to create the population in the pool.  Parents are spliced, with a note pitch or rhythm single point mutation probability of 0.1.  The mutation rate is set to this relatively high value as the splicing and recombination manipulations of themes occurs through the operation of melody agents, meaning mutation is the only method of adding new melodic content to the generated theme.

\subsubsection{Percussive Rhythm}

A simple percussive rhythm agent based on the RNN model used by Hutchings \cite{hutchings2017talking} was implemented with the lowest melody agent  rhythmically doubled by the lowest percussive instrument and fed into the neural network to produce the other percussive parts in the specified style.

\section{Implementation in video games}

To test the effectiveness of the AMS, for the intended task of real-time scoring of game-world events in video games, the model was implemented in two games: \emph{Zelda Mystery of Solarus} (MoS) \cite{game:zelda} and \emph{StarCraft II} \cite{game:starcraft}.  The games were selected for having different mechanics, visual styles and settings, as well as having open source code or exposed game-states that could be used to model game-world contents.  The activation of objects and environments (Fig.~\ref{fig:visualisation}), through explicit programmatic rules, spreads to affect categories, which in turn affects parameters of the musical compositions. 

\textcolor{red}{A} list of key concepts that players would be likely to engage with over the first thirty minutes of gameplay was established for each game and activations for concepts and emotions were calibrated.  In Zelda: MoS, concepts such as `Grandma' and `Bakery'  were given an activation of 100 when on screen, because there was only one of each of these object types in the game.  Other objects such as `Heart' and `Big Green Knight' were given activation levels of 20 for appearing on screen and increased by 10 for each interaction the player character had with them.  Attacking a knight, or being attacked by knight counted as an interaction.

For StarCraft II, threat was set by the proportion of enemy units in view compared to friendly units.  The rate of wealth growth was chosen as an activator of happiness with 5 activation points added for each 100 units of resources mined.  Sadness activation increased 5 points by death of friendly units, tenderness activation increased 5 points by units performing healing functions and excitement activation increased 10 points by the number of enemy units visible.  Three environment nodes were created: the player's base, enemy base and wasteland in-between.

Participants were invited to participate in the study on public social media pages for students at Monash University, gamers and game developers in Melbourne, Australia and divided into two groups: Group A and Group B, \textcolor{red}{with seventeen participants in each group.}  Experiments were conducted in an isolated, quiet room and questionnaires were completed unobserved and anonymously.  \textcolor{red}{The reasoning behind this study design was to avoid experience conflation by having the same game played with different music conditions.} Group A played Zelda: MoS with its original score and StarCraft with a score generated with the AMS.  Group B played Zelda: MoS with the AMS and StarCraft with the original score.  The order that the games were presented in alternated for each participant and participants were not given any information about the music condition or system in use.

Players would play their first game for twenty minutes.  After twenty minutes the game would automatically stop and the first game questionnaire would appear.  Participants were asked if they had played the game before, to describe the game and music with words, if they notice the mood of the music change during gameplay and to list any game related associations they have with a played music theme.  

\textcolor{red}{One question of the questionnaire included a 10 second audio clip that the participant would use to report any  game-world contents that they associated with that music.  Before the study began a list of key words was defined (see Table \ref{correctAssociation}) for each condition to classify a `correct' association, meaning an association between music and content that had co-occurrence during gameplay or content that was known to trigger the music.}  For Starcraft, a recording of the original score used during a conflict-free segment of gameplay was used for participants in group B and the composer written theme for `SCV' units played for participants in group A.  For Zelda: MoS, a recording of the original score used in outdoor areas was played to participants in group B and the composer written theme for `Sword' was played to participants in group A.

\begin{table}
\centering
\caption{Accepted terms for established concepts}
\setlength\tabcolsep{3.2pt}
\label{correctAssociation}
\begin{tabular}{@{}lll@{}}
\toprule
                 & Associated concept & Accepted terms                        \\ \midrule
Zelda      & Outdoors  & town, village, outdoors, outside      \\
Zelda AMS   & Sword                      & sword, fighting, battle               \\
Starcraft     & No conflict                & building, constructing, mining, peace \\
Starcraft AMS & SCV                        & SCV, building, constructing, mining   \\ \bottomrule
\end{tabular}
\end{table}

Participants were then asked to provide ratings of immersion, correlation of music with actions and questions related to the quality of the music itself for the gaming session with each game.  After submitting answers participants were invited to take a rest before completing another game session and questionnaire with the other game.

\subsection{Results}

Example videos of each game being played with music generated by the AMS are provided for reference.\footnote{\url{https://monash.figshare.com/s/1d863d9aa90ca4a97aab}}

Most of the participants were unfamiliar with the games used in the study.  For Starcraft II, 17\% of users reported having played the game before and only one participant, representing 3\% of participants had played Zelda: MoS.

\begin{table}
\centering
\caption{Mann-Whitney U analysis of difference in reported immersion and correlation of events and music between original game score and AMS gameplay conditions.}
\setlength\tabcolsep{3.2pt}
\label{resultstable}
\begin{tabular}{@{}llllllll@{}}
\toprule
            & \multicolumn{2}{l}{Immersion}            &             &\hphantom{pad} & \multicolumn{2}{l}{Correlation}        &      \\ \midrule
            & z         & p         & $\eta^{2}$    & & z         & p         & $\eta^{2}$    \\ \midrule
Zelda       & -1.17     & 0.121     & 0.042       & & -1.58     & 0.057      & 0.075     \\
Starcraft   & -1.62     & 0.053     & 0.079       & & -1.60     & 0.055      & 0.077     \\ \bottomrule
\end{tabular}
\end{table}

\textcolor{red}{A small positive effect (Table \ref{resultstable}) was observed in both reported immersion and correlation of music with events in AMS conditions compared to original score conditions for the same game, for both games.}

\textcolor{red}{Using a Wilcoxon Signed-rank non-directional test, a significant positive increase in rank sum ($p < 0.05$) was observed for both groups in reporting perceived effect of music on immersion when AMS was used, independent of game choice.}  Further, in no cases did participants report that AMS generated music `greatly' or `somewhat' reduced immersion.

\textcolor{red}{Despite the increase in self-reported correlation of events and music, fewer `correct' concept terms were listed for the short audio samples in the questionnaire in AMS conditions for both games.  For Zelda correct terms were reported by 70\% of participants in original and 53\% in AMS conditions.  For Starcraft this rate was 59\% for original and 41\% for AMS.}

\subsection{Discussion}
\textcolor{red}{The study involved short interactions with a game that does not represent the extendend periods of time that many people play games for.  This short period also limits the complexity of the spreading activation model that can be created and personalisation of the music experience that such complexity would bring.}

\textcolor{red}{However, the study shows that the AMS can be successfully implemented in games and that self-reported immersion and correlation of events with music is increased using the AMS for short periods.  This result supports the use of the spreading activation model to combine object, emotion and environment concepts.  Listening to the music tracks it becomes apparent that the overall music quality of the AMS is not that of a skilled composer, but this goal is outside the scope of this project.  Musical quality could be enhanced through improvements to the composition techniques within the framework presented.}

\section{Conclusions}

In this paper we have documented a process of identifying key issues in scoring dynamic musical content in video games, finding suitable emotion models and implementing them in a video game scoring system.  The journey through consideration of alternate modelling approaches, exploratory study, novel listener study and modelling approach, implementation and testing in games demonstrates the multitude of academic and pragmatic concerns developers can expect to face in producing new emotion-driven, adaptive scoring systems.   

By focusing on the criteria of the emotion model for the specific application, additional development was required in establishing a novel listener study design, but greater confidence in the suitability of the model was gained and positive real-world implementation results were achieved.  The field of affective computer music could benefit from further exploration of designed-for-purpose data collection techniques with regards to emotion.

The results of the game implementation study suggest that using affect categories within a knowledge activation model as a communicative layer between game-world events and the music composition system can lead to improved immersion and perceived correlation of events with music for gamers, compared to current standard video-game scoring approaches.

\textcolor{red}{We argue for the need of a fully integrated, end-to-end approach to adaptive game music that sees the music engine integrated at an early stage of development and built on models of cognition, music psychology and narrative.  This approach has a broad and deep scope. As research in this area continues, expert knowledge and pragmatic design decisions will continue to be needed to produce actual music systems from models and theoretical frameworks.  Individual design decisions in the implementation presented in this paper, including the choice of generative models and calibration of parameters, raise individual research questions that can be pursued within the general framework presented or in isolation.}

\subsection{Future Work}

We intend to undertake further listener and gamer studies to establish more generalised correlation between music and emotion. These studies support developing more advanced adaptive music systems.  The listener study design presented in this paper gives us a suitable platform for collecting further data for a larger range of music compositions.  Positive results from the gamer survey based on modifying existing games not designed for the AMS encourage ground-up implementations in new games for tighter integration with the mechanics of specific gameplay.

\textcolor{red}{The AMS presented assumes that knowledge of upcoming events is not present.  While this scenario may be the case in some open-world games like \textit{Minecraft}, many games have written narratives that can provide this knowledge \emph{a priori}.  Work by Scirea et al. has demonstrated benefits of musical foreshadowing in games \cite{scirea2014evaluating}, an important addition that could be implemented in this AMS.}

\section*{Acknowledgements}
This research was supported by an Australian Government Research Training Program (RTP) Scholarship.

\newpage
\appendix
\setcounter{table}{0}
\renewcommand{\thetable}{A\arabic{table}}

\begin{table}[h]
\caption{Listener study rank of tracks by emotion}
\label{tab:emotionTable}
\begin{tabularx}{0.47\textwidth}{XXXXXXX}
\toprule
Track & Hap. & Exc. & Ang. & Sad. & Ten. & Thr. \\ \midrule
1     & 25        & 22         & 7     & 1       & 15         & 4      \\
2     & 7         & 6          & 25    & 28      & 14         & 26     \\
3     & 8         & 25         & 26    & 11      & 1          & 27     \\
4     & 4         & 10         & 10    & 27      & 26         & 3      \\
5     & 9         & 14         & 17    & 10      & 5          & 23     \\ 
6     & 14        & 5          & 1     & 29      & 29         & 14     \\
7     & 15        & 7          & 22    & 23      & 18         & 13     \\
8     & 2         & 11         & 30    & 20      & 3          & 28     \\
9     & 13        & 16         & 8     & 8       & 22         & 20     \\
10    & 17        & 9          & 5     & 25      & 25         & 1      \\ \midrule
11    & 22        & 21         & 20    & 5       & 7          & 24     \\
12    & 18        & 23         & 19    & 2       & 4          & 17     \\
13    & 29        & 20         & 16    & 4       & 8          & 21     \\
14    & 6         & 27         & 12    & 15      & 11         & 19     \\
15    & 16        & 12         & 27    & 19      & 16         & 9      \\ 
16    & 28        & 30         & 21    & 17      & 6          & 29     \\
17    & 5         & 3          & 6     & 30      & 28         & 15     \\
18    & 24        & 28         & 11    & 9       & 20         & 16     \\
19    & 11        & 15         & 18    & 21      & 21         & 12     \\
20    & 20        & 18         & 4     & 18      & 24         & 6      \\ \midrule
21    & 10        & 4          & 28    & 22      & 9          & 30     \\
22    & 26        & 19         & 3     & 16      & 19         & 2      \\
23    & 19        & 17         & 15    & 3       & 13         & 7      \\
24    & 21        & 1          & 14    & 26      & 30         & 10     \\
25    & 30        & 13         & 24    & 7       & 2          & 22     \\
26    & 27        & 2          & 2     & 13      & 27         & 5      \\
27    & 1         & 8          & 9     & 24      & 23         & 11     \\
28    & 23        & 26         & 13    & 6       & 10         & 18     \\
29    & 3         & 24         & 29    & 14      & 17         & 25     \\
30    & 12        & 29         & 23    & 12      & 12         & 8      \\ \bottomrule
\end{tabularx}
\end{table}

\begin{table}[!h]
\centering
\caption{Pearson's correlation coefficient (r) of music pitch properties by emotion category.  Correlations with p $<$ 0.01 in bold. M = Mean, SD = Standard deviation, ST = Semi-tones}
\label{tab:pitchEmotion}
\setlength\tabcolsep{3.2pt}
\begin{tabularx}{0.5\textwidth}{XXXXXXXX}
\toprule
           & Phrase length (M) & Phrase length (SD) & Pitch range (ST) & Pitch interval (M) & Diatonic (\%) & Notes/s & Note velocity (M) \\ \midrule
Ang.      & 0.032             & 0.249        & 0.200       & 0.326              & -0.046 & 0.270           & \textbf{0.720} \\
Exc. & 0.256             & 0.184        & -0.110      & 0.044              & 0.074  & \textbf{0.710}  & 0.064 \\
Hap.  & -0.126            & -0.123       & -0.317      & -0.113             & \textbf{0.665} & 0.349           & 0.001 \\
Sad.    & -0.195            & -0.190       & 0.229       & -0.091             & -0.430 & \textbf{-0.779} & -0.121  \\
Ten. & -0.143            & -0.378       & -0.011      & -0.240             & 0.016 & \textbf{-0.639} & -0.465  \\
Thr.     & 0.105             & 0.267        & 0.239       & \textbf{0.578}     & -0.191 & 0.214           & \textbf{0.622}  \\ \bottomrule
\end{tabularx}
\end{table}

\ifCLASSOPTIONcaptionsoff
  \newpage
\fi


\bibliographystyle{IEEEtran}

\bibliography{IEEEabrv,bibtex/bib/IEEEexample}
%

\end{document}